\documentclass[floatfix,aps,prl,twocolumn,showpacs,superscriptaddress]{revtex4}

\usepackage{graphicx}
\usepackage[english]{babel}
\usepackage{bm}    

\def\half{ {\frac{1}{2}} }

\def\reff#1{(\ref{#1})}
\newcommand{\be}{\begin{equation}}
\newcommand{\ee}{\end{equation}}
\newcommand{\<}{\langle}
\renewcommand{\>}{\rangle}

\def\spose#1{\hbox to 0pt{#1\hss}}
\def\ltapprox{\mathrel{\spose{\lower 3pt\hbox{$\mathchar"218$}}
 \raise 2.0pt\hbox{$\mathchar"13C$}}}
\def\gtapprox{\mathrel{\spose{\lower 3pt\hbox{$\mathchar"218$}}
 \raise 2.0pt\hbox{$\mathchar"13E$}}}

\newcommand{\scrc}{{\cal C}}

\newcommand{\scre}{{\cal E}}

\newcommand{\scrl}{{\cal L}}

\newcommand{\scrn}{{\cal N}}
\newcommand{\scro}{{\cal O}}

\newcommand{\scrs}{{\cal S}}

\begin{document}

\title{Dynamic critical behavior of the Chayes--Machta--Swendsen--Wang
       algorithm}

\author{Youjin Deng}
\affiliation{Department of Physics, New York University,
      4 Washington Place, New York, NY 10003, USA}
\author{Timothy M.~Garoni}
\affiliation{Department of Physics, New York University,
      4 Washington Place, New York, NY 10003, USA}
\author{Jonathan Machta}
\affiliation{Department of Physics, University of Massachusetts,
      Amherst, MA 01003, USA}
\author{Giovanni Ossola}
\affiliation{Institute of Nuclear Physics, NCSR ``Demokritos'',
      15310 Athens, Greece}
\author{Marco Polin}
\affiliation{Department of Physics, New York University,
      4 Washington Place, New York, NY 10003, USA}
\author{Alan D. Sokal}
\affiliation{Department of Physics, New York University,
      4 Washington Place, New York, NY 10003, USA}
\affiliation{Department of Mathematics,
      University College London, London WC1E 6BT, UK}

\date{May 17, 2007}

\begin{abstract}
We study the dynamic critical behavior
of the Chayes--Machta dynamics for the
Fortuin--Kasteleyn random-cluster model,
which generalizes the Swendsen--Wang dynamics for the $q$-state Potts model
to noninteger $q$,
in two and three spatial dimensions, by Monte Carlo simulation.
We show that the Li--Sokal bound $z \ge \alpha/\nu$
is close to but probably not sharp in $d=2$,
and is far from sharp in $d=3$, for all $q$.
The conjecture $z \ge \beta/\nu$ is false (for some values of $q$)
in both $d=2$ and $d=3$.
\end{abstract}



\maketitle


Monte Carlo simulations in statistical mechanics \cite{Binder_79-92}
and quantum field theory \cite{Monte_Carlo_QFT}
typically suffer from {\em critical slowing-down}\/
\cite{Hohenberg_77,Sokal_Cargese_96}:
the autocorrelation (relaxation) time $\tau$ diverges
as the critical point is approached,
most often like $\tau \sim \xi^z$,
where $\xi$ is the spatial correlation length
and $z$ is a dynamic critical exponent.
For local algorithms, one usually has $z \approx 2$.
This effect severely limits the efficiency of Monte Carlo studies
of critical phenomena in statistical mechanics
and of the continuum limit in quantum field theory.

An important advance was made in 1987 with the invention of
the Swendsen--Wang (SW) cluster algorithm \cite{Swendsen_87}
for simulating the $q$-state ferromagnetic Potts model \cite{Potts_52,Wu_82+84}
at positive integer $q$.
The SW algorithm is based on passing back and forth
between the Potts spin representation
and the Fortuin--Kasteleyn (FK) bond representation \cite{FK_69+72,Edwards_88}.
This algorithm does not eliminate critical slowing-down,
but it radically reduces it compared to local algorithms.
Much effort has therefore been devoted,
for both theoretical and practical reasons,
to understanding the dynamic critical behavior of the SW algorithm
as a function of the spatial dimension $d$
and the number $q$ of Potts spin states \cite{Ossola-Sokal}.
Unfortunately, it is very difficult to develop a physical understanding
from the small number of ``data points'' at our disposal:
second-order transitions occur only for
$(d,q) = (2,2)$, (2,3), (2,4), (3,2) and (4,2) \cite{note_5dising}.

A further advance was made in 1998
by Chayes and Machta (CM) \cite{Chayes-Machta},
who devised a cluster algorithm for simulating
the FK random-cluster model \cite{FK_69+72,Grimmett_06} ---
which provides a natural extension of the Potts model to noninteger $q$ ---
at any real $q \ge 1$.
The CM algorithm generalizes the SW algorithm
and in fact reduces to (a slight variant of) it when $q$ is an integer.
By using the CM algorithm, we can study the dynamic critical behavior
of the SW--CM dynamic universality class
as a function of the {\em continuous}\/ variable $q$
throughout the range $1 \le q \le q_c(\scrl)$,
where $q_c(\scrl)$ is the maximum $q$ for which the transition is
second-order on the lattice $\scrl$ \cite{fn_1st2nd}.
This vastly enhances our ability to make theoretical sense
of the numerical results.

In this Letter we report detailed measurements
of the dynamic critical behavior of the CM algorithm
for two-dimensional random-cluster models with $1 \le q \le 4$ \cite{cm_2d}
and for three-dimensional models with $q = 1.5,\, 1.8,\, 2,\, 2.2$ \cite{cm_3d}.
Among other things, we find strong evidence {\em against}\/
the conjecture $z \ge \beta/\nu$
recently proposed by two of us \cite{Ossola-Sokal},
which had seemed plausible from the data for integer $q$.

The {\em FK random-cluster model}\/ with parameter $q > 0$
is defined on any finite graph $G=(V,E)$ by the partition function
\begin{equation}
   Z \;=\;  \sum_{A \subseteq E} q^{k(A)} \prod_{e \in A} v_e
   \;,
 \label{eq.ZRC}
\end{equation}
where $A$ is the set of ``occupied bonds''
and $k(A)$ is the number of connected components (``clusters'')
in the graph $(V,A)$;
here $\{ v_e \}$ are nonnegative edge weights.
For $q=1$ this reduces to independent bond percolation \cite{Stauffer_92}
with occupation probabilities $p_e = v_e/(1+v_e)$;
for integer $q \ge 1$ it provides a graphical representation
of the $q$-state ferromagnetic Potts model
with nearest-neighbor couplings $\{J_e\}$, where $v_e = e^{\beta J_e} - 1$.

It is convenient to consider a {\em generalized random-cluster (RC) model}\/
\cite{cm_3d,clu_loop_prl}
\be
   Z  \;=\;
   \sum_{A \subseteq E}
       \left( \prod_{e \in A} v_e \!\right)
       \!
       \left( \prod_{i=1}^k W(H_i) \!\right)
       \,,
 \label{def.genRC}
\ee
where $H_1,\ldots,H_k$ are the connected components of the graph $(V,A)$,
and $\{ W(H) \}$ are nonnegative weights associated to the connected subgraphs
$H$ of $G$.
The model \reff{def.genRC} reduces to the FK model \reff{eq.ZRC}
if $W(H) = q$ for all $H$;
other special cases include an FK representation
for the Potts model in a magnetic field \cite{FK_Potts_magfield}
and various loop models \cite{clu_loop_prl}.

Now let $m$ be a positive integer, and let us decompose each weight $W(H)$
into $m$ nonnegative pieces, any way we like:
$W(H) = \sum_{\alpha=1}^m W_\alpha(H)$.
The first step of our generalized Chayes--Machta algorithm,
given a bond configuration $A$, is to choose, independently
for each connected component $H_i$, a ``color'' $\alpha \in \{1,\ldots,m\}$
with probabilities $W_\alpha(H_i)/W(H_i)$;
this color is then assigned to all the vertices of $H_i$.
The vertex set $V$ is thus partitioned as $V = \bigcup_{\alpha=1}^m V_\alpha$.
It is not hard to see that, conditioning on this decomposition,
the bond configuration is nothing other than a
generalized RC model with weights $\{ W_\alpha(H) \}$
on the induced subgraph $G[V_\alpha]$, independently for each $\alpha$.

We now have the right to update these generalized RC models
by any valid Monte Carlo algorithm.
One valid update is ``do nothing'';
this corresponds to the ``inactive'' colors
of Chayes and Machta \cite{Chayes-Machta}.
Of course, we must also include at least one nontrivial update.
The basic idea is to have at least one color
for which the weights $W_\alpha(H)$ are ``easy'' to simulate.
In particular, when $W(H) = q$ for all $H$
(the standard FK random-cluster model),
we can take $W_\alpha(H) = 1$ for one or more colors $\alpha$
(the so-called ``active'' colors);
the corresponding model on $G[V_\alpha]$ is then independent bond percolation,
which can be trivially updated.
Since we must have $W_\alpha(H) \le W(H)$, this works whenever $q \ge 1$.
More generally, if $q \ge k$, then we can have $k$ active colors.
If $q$ is an integer and we take $k=q$, we recover the standard SW algorithm.

We used the CM algorithm to simulate the random-cluster model
in dimensions $d=2,3$ on hypercubic lattices of size $L^d$
with periodic boundary conditions.
We measured the ``energy-like'' observables
$\scrn = \#$ of occupied bonds and
$\scre' = \#$ of nearest-neighbor pairs belonging to the same cluster;
the cluster-size moments $\scrs_m = \sum |C|^m$ for various values of $m$,
where $|C|$ is the number of sites in the cluster $C$;
and the size $\scrc_i$ of the $i$th-largest cluster for $i=1,2,3$.

For any observable $\scro$,
let $\rho_{\scro} (t)$ be its normalized autocorrelation function.
Then define the exponential autocorrelation time
\begin{equation}
   \tau_{{\rm exp},\scro} \;=\;
   \limsup_{t \to \pm\infty} {|t| \over   - \log |\rho_{\scro}(t)|}
\end{equation}
and the integrated autocorrelation time
\begin{equation}
   \tau_{{\rm int},\scro}   \;=\;
   \half \sum_{t = -\infty}^{\infty}  \rho_{\scro}(t)
   \;.
\end{equation}
Typically all observables $\scro$
(except those that, for symmetry reasons,
 are ``orthogonal'' to the slowest mode)
have the same value $\tau_{{\rm exp},\scro} = \tau_{{\rm exp}}$.
However, they may have very different amplitudes of ``overlap''
with this slowest mode;
in particular, they may have very different values of the
integrated autocorrelation time,
which controls the efficiency of Monte Carlo simulations
\cite{Sokal_Cargese_96}.
We define dynamic critical exponents $z_{\rm exp}$
and $z_{{\rm int},\scro}$ by
$\tau_{\rm exp} \sim \xi^{z_{\rm exp}}$
and $\tau_{{\rm int},\scro} \sim \xi^{z_{{\rm int},\scro}}$.
On a finite lattice at criticality, $\xi$ can here be replaced by $L$.

%
%

\begin{figure}[h]
\begin{center}
\includegraphics[width=1.02\columnwidth]{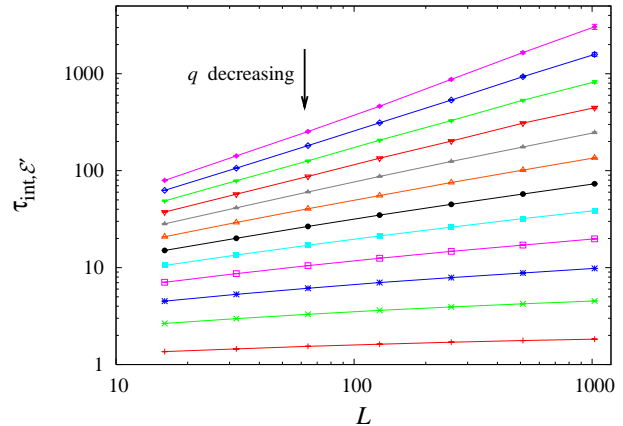}
\end{center}
\vspace*{-8mm}
\caption{
   Integrated autocorrelation times $\tau_{{\rm int},\scre'}$
   versus lattice size $L$ for the critical two-dimensional
   random-cluster model at $1.25 \le q \le 4$,
   simulated using the Chayes--Machta algorithm with $k=1$.
}
\label{fig1}
\end{figure}


%
%


\begin{table}[h]
\begin{center}
\begin{tabular}{|ccccrc|}
\hline
   $q$   &  Fit  &  \multicolumn{1}{c}{$L_{\rm min}$}  &
        $z_{{\rm int},\scre'}$  &
        \multicolumn{1}{c}{$\alpha/\nu$}  &  $\beta/\nu$  \\
\hline
   1.00  &    exact     & ---  &   0         &  $-0.5000$   &   0.1042    \\
   1.25  &  $A+BL^{-p}$ & 128  &   0         &  $-0.3553$   &   0.1112    \\
   1.50  &  $A+BL^{-p}$ &  32  &   0         &  $-0.2266$   &   0.1168    \\
   1.75  &  $AL^z + B$  &  16  &   0.06(1)   &  $-0.1093$   &   0.1213    \\
   2.00  &  $AL^z + B$  &  32  &   0.14(1)   &  0 (log)     &   0.1250    \\
   2.25  &  $AL^z + B$  &  32  &   0.24(1)   &  0.1036      &   0.1280    \\
   2.50  &  $AL^z + B$  &  32  &   0.31(1)   &  0.2036      &   0.1303    \\
   2.75  &  $AL^z + B$  &  16  &   0.40(2)   &  0.3017      &   0.1321    \\
   3.00  &  $AL^z + B$  &  32  &   0.49(1)   &  0.4000      &   0.1333    \\
   3.25  &  $AL^z + B$  &  64  &   0.57(1)   &  0.5013      &   0.1339    \\
   3.50  &  $AL^z$      &  16  &   0.69(1)   &  0.6101      &   0.1338    \\
   3.75  &  $AL^z$      &  32  &   0.78(1)   &  0.7376      &   0.1324    \\
   4.00  &  $AL^z + B$  &  32  &   0.93(2)   &  1.0000      &   0.1250    \\
\hline
\end{tabular}
\end{center}
\vspace*{-4mm}
\caption{
   Dynamic critical exponents $z_{{\rm int},\scre'}$
   for two-dimensional random-cluster model
   as a function of $q$,
   with preferred fit and minimum $L$ value used in the fit.
   Error bars are one standard deviation, statistical error only.
   The exact exponents $\alpha/\nu$ and $\beta/\nu$
   are shown for comparison \cite{note_exact_2d_exponents}.
}
\label{table1}
\end{table}


We began by performing simulations on the square lattice ($d=2$)
at the exact critical point $v_c(q) = \sqrt{q}$ \cite{Baxter_book}
for $1.25 \le q \le 4$ in steps of 0.25
and lattice sizes $16 \le L \le 1024$,
using all positive integer values of $k \le q$.
We estimated the integrated autocorrelation times
$\tau_{{\rm int},\scro}$ by the automatic windowing method
described in \cite{Madras_88,Ossola-Sokal}.
The complete set of runs used approximately 14.8~yr CPU time
on a 1266 MHz Pentium III Tualatin processor.

The autocorrelation functions of $\scrn$, $\scre'$ and $\scrs_m$
are in all cases very close to a pure exponential.
In Fig.~\ref{fig1} we plot $\tau_{{\rm int},\scre'}$ (for $k=1$) versus $L$,
and in Table~\ref{table1} we report the estimated dynamic critical exponents
$z_{{\rm int},\scre'}$.
Our data also show that, as expected, the exponents are independent of $k$,
and we have roughly $\tau \propto 1/k$.

%
%

\begin{figure}[t]
\begin{center}
\includegraphics[width=1.02\columnwidth]{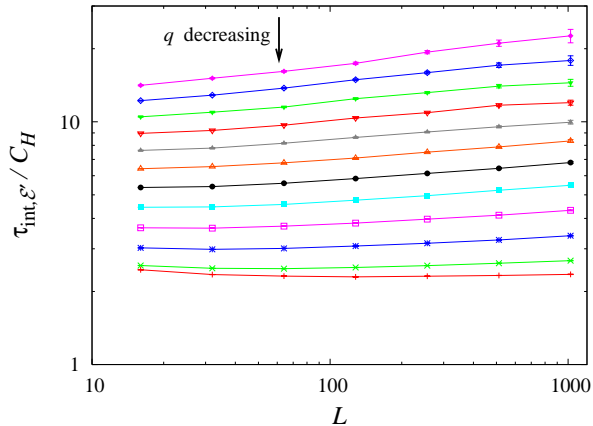}
\end{center}
\vspace*{-8mm}
\caption{
   Integrated autocorrelation times $\tau_{{\rm int},\scre'}$
   divided by specific heat $C_H$, versus lattice size $L$,
   for the critical two-dimensional random-cluster model
   at $1.25 \le q \le 4$,
   simulated using the Chayes--Machta algorithm with $k=1$.
}
\label{fig2}
\end{figure}


Since the Li--Sokal bound
$\tau_{{\rm int},\scrn}, \tau_{{\rm int},\scre'} \ge {\rm const} \times C_H$
and hence
$z_{{\rm int},\scrn}, z_{{\rm int},\scre'} \ge \alpha/\nu$,
originally proven \cite{Li-Sokal} for the Swendsen--Wang algorithm,
can also be proven \cite{cm_3d} for the Chayes--Machta algorithm
(at least for $\scrn$),
it is of interest to analyze its possible sharpness \cite{note_q=4}.
In Fig.~\ref{fig2} we plot $\tau_{{\rm int},\scre'}/C_H$ versus $L$,
in an attempt to determine whether this ratio
is bounded or not as $L \to\infty$.
The results are far from clear,
but our best guess is that $\tau_{{\rm int},\scre'}/C_H$
diverges as $L \to\infty$, either as a small power or as a logarithm.
However, the precise behavior needs to be explored by
simulations at larger $L$.

%

On the other hand, Ossola and Sokal \cite{Ossola-Sokal} recently conjectured,
on the basis of the ``data points''
$(d,q) = (2,2)$, (2,3), (2,4), (3,2) and (4,2),
that $z_{{\rm int},\scre'} \ge \beta/\nu$;
and they even speculated that we might have the {\em equality}\/
$z_{{\rm int},\scre'} = \max(\alpha/\nu,\beta/\nu)$.
%
%
%
The data for noninteger $q$ now shed light on this conjecture:
for $q=1.25,\, 1.5$ there is modest evidence
(and for $q=1.75$ there is weak evidence)
that $z_{{\rm int},\scre'} < \beta/\nu$, i.e.\ that even the weak form
of the Ossola--Sokal conjecture is {\em false}\/.

We next performed simulations on the simple-cubic lattice ($d=3$)
for $q=1.5,\, 1.8,\, 2.2$
(see also \cite{Ossola-Sokal} for $q=2$)
and lattice sizes $4 \le L \le 256$,
using $k = $ the largest integer $\le q$.
We located the critical point by a finite-size-scaling analysis
using the ratio $R = \<\scrs_4\> / \<\scrs_2^2\>$,
as in \cite{forests_3d4d_prl}.
The complete set of runs
used approximately 21.5~yr CPU time on a 3.2 GHz Xeon EM64T processor.

%
%

\begin{figure}[t]
\begin{center}
\includegraphics[width=1.02\columnwidth]{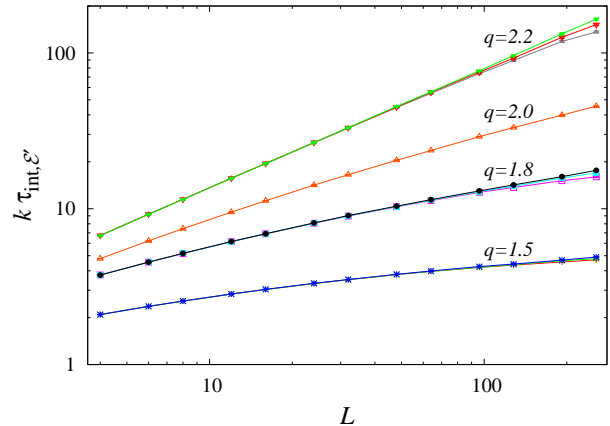}
\end{center}
\vspace*{-8mm}
\caption{
   $k \,\tau_{{\rm int},\scre'}$ versus lattice size $L$
   for Chayes--Machta simulations of the three-dimensional
   random-cluster model with $q=1.5,\, 1.8,\, 2.2$
   at three near-critical temperatures,
   taking $k=\lfloor q \rfloor$. 
   Data for $q=2$, $k=2$ at approximate criticality
   are from \cite{Ossola-Sokal}.
}
\label{fig3}
\end{figure}


%
%

\begin{table}[t]
\vspace*{2mm} \hspace*{-0cm} 
\begin{center}
\begin{tabular}{|cccccc|}
\hline
   $q$  &  Fit   & \multicolumn{1}{c}{$L_{\rm min}$} &
                   $z_{{\rm int},\scre'}$  &  $\alpha/\nu$  & $\beta/\nu$ \\
\hline
   1.5  &  $AL^z$ &  96  &   0.13(1)   &  $-0.32(4)$  &  0.500(4)    \\
   1.8  &  $AL^z$ &  96  &   0.29(1)   &  $-0.15(5)$  &  0.5117(6)   \\
    2   &  $AL^z$ &  96  &   0.46(3)   &  0.174(1)    &  0.5184(1)   \\
   2.2  &  $AL^z$ &  24  &   0.76(1)   &  0.50(4)     &  0.508(4)  \\
\hline
\end{tabular}
\end{center}
\vspace*{-4mm}
\caption{
   Dynamic critical exponents $z_{{\rm int},\scre'}$
   and static exponents $\alpha/\nu$ and $\beta/\nu$
   for three-dimensional random-cluster model.
   For $q=2$, dynamic data are from \cite{Ossola-Sokal}
   and static exponents are from \cite{d=3_Ising}.
}
\label{table2}
\end{table}

\begin{figure}[t]
\begin{center}
\includegraphics[width=1.02\columnwidth]{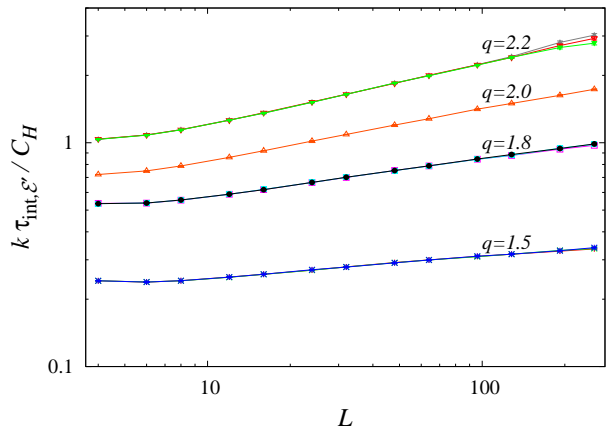}
\end{center}
\vspace*{-8mm}
\caption{
   $k \,\tau_{{\rm int},\scre'}/C_H$ versus lattice size $L$
   for Chayes--Machta simulations of the near-critical three-dimensional
   random-cluster model at $q=1.5,\, 1.8,\, 2,\, 2.2$,
   with $k=\lfloor q \rfloor$.
   Data for $q=2$ at approximate criticality are from \cite{Ossola-Sokal}.
}
\label{fig4}
\end{figure}

The autocorrelation functions of $\scrn$, $\scre'$ and $\scrs_m$
are again very close to a pure exponential.
In Fig.~\ref{fig3} we plot $k \tau_{{\rm int},\scre'}$ versus $L$
(multiplying by $k$ makes the results for different $q$ comparable)
for three temperatures very near criticality.
In Table~\ref{table2} we report the estimated
dynamic critical exponents $z_{{\rm int},\scre'}$
and static critical exponents $\alpha/\nu$ and $\beta/\nu$.
In Fig.~\ref{fig4} we plot $k\tau_{{\rm int},\scre'}/C_H$ versus $L$.
It seems clear that, for all four values of $q$,
the Li--Sokal bound is far from sharp.
On the other hand, from Table \ref{table2} it seems clear that
for $q=1.5,\, 1.8$ we have the strict inequality
$z_{{\rm int},\scre'} < \beta/\nu$,
once again ruling out the Ossola--Sokal conjecture even in its weak form.

%
%

The dynamic critical behavior of the SW--CM
dynamic universality class in dimension $d=3$ therefore remains a mystery.
Clearly, some new {\em physical}\/ principle,
beyond the slow equilibration of the energy
embodied in the Li--Sokal bound \cite{Li-Sokal},
needs to be discovered.

One clue might be provided by our analysis \cite{cm_3d}
of the CM algorithm on the complete graph (mean-field limit),
generalizing the analysis in \cite{PeBeKaDo96} of the SW algorithm.
Taking $k=1$ and defining a ``magnetization'' $m$
to be the fraction of sites in the largest cluster,
we obtain for $1 \le q \le 2$ the approximate difference equation
(generalizing \cite[eq.~(10)]{PeBeKaDo96})
\begin{equation}
   m^\prime \,=\,
   \frac{2q-2}{q}m - \frac{4t}{q^2} + \frac{8(q-1)tm}{3q^2}
     - \frac{2(q-1)^2 m^2}{3q}
   \;
\end{equation}
where $m'$ is the value of $m$ after a sweep in which the active group
contains the largest cluster,
and $t$ is the deviation from the critical temperature.
Clearly $q=2$ is a special case because the coefficient
of the linear term equals 1:  we have $\beta=1/2$ and $z=1$,
and it is clear from the derivations \cite{PeBeKaDo96,cm_3d}
that $z$ is actually $\beta/\nu$.
For $1 \leq q<2$, by contrast, both the statics and dynamics are in the
percolation universality class with $\beta=1$ and $z=0$:
small perturbations from equilibrium relax exponentially
with a finite autocorrelation time
$\tau_{{\rm exp},m}= q/\log[q/(2q-2)]$ that diverges as $q \uparrow 2$.
We conjecture that a similar behavior holds
above the upper critical dimension,
which for $q < 2$ is presumably $d=6$.
Our numerical data \cite{cm_3d} confirm the behavior $z=0$ for $1 \leq q<2$
with $\tau \propto 1/(2-q)$ as $q \uparrow 2$,
but not the predicted amplitude.

Details of these simulations and their data analysis
will be reported separately \cite{cm_2d,cm_3d}.

\begin{acknowledgments}
This work was supported in part by NSF grants PHY--0116590 and PHY--0424082.
\end{acknowledgments}



\begin{thebibliography}{99}
\bibitem{note_5dising}  Second-order transitions occur also for
   the Ising ($q=2$) model in dimensions $d > 4$,
   but here the static behavior is mean-field.
   One expects the dynamic critical exponents likewise to be
   dimension-independent for $d \ge 4$
   (with possible multiplicative logarithmic corrections at $d=4$).
\end{thebibliography}
\end{document}